\documentclass[12pt]{iopart}

\usepackage{siunitx}
\usepackage{upgreek}
\usepackage{xspace}
\usepackage{graphicx}
\usepackage{subfig}
\usepackage{textalpha}
\usepackage{booktabs}
\usepackage{multirow}
\usepackage{biblatex}
\usepackage[
	colorlinks=true,
	urlcolor=blue,
	anchorcolor=blue,
	citecolor=blue,
	filecolor=blue,
	linkcolor=blue,
	menucolor=blue,
	pagecolor=blue,
	linktocpage=true,
	pdfa=true
]{hyperref}

\DeclareSIUnit{\keV}{\kilo\electronvolt}

\newcommand{\rbkr}{\textsuperscript{83}Rb\slash\textsuperscript{83m}Kr\xspace}
\newcommand{\rb}{\textsuperscript{83}Rb\xspace}
\newcommand{\kr}{\textsuperscript{83m}Kr\xspace}
\newcommand{\kalpha}{K\textsubscript{\textalpha}}
\newcommand{\kbeta}{K\textsubscript{\textbeta}}
\newcommand{\krline}[3]{#1\textsubscript{#2}-#3{}}

\begin{document}

\title[Characterization of Silicon Drift Detectors with Electrons for the TRISTAN Project]{Characterization of Silicon Drift Detectors with Electrons for the TRISTAN Project}

\author{S~Mertens$^{1,2}$\footnote{Corresponding author}, T~Brunst$^{1,2}$, M~Korzeczek$^3$, M~Lebert$^{1,2}$, D~Siegmann$^{1,2}$, A~Alborini$^4$, K~Altenmüller$^1$, M~Biassoni$^5$, L~Bombelli$^4$, M~Carminati$^{6,7}$, M~Descher$^3$, D~Fink$^2$, C~Fiorini$^{6,7}$, C~Forstner$^{1,2}$, M~Gugiatti$^{6,7}$, T~Houdy$^{1,2}$, A~Huber$^3$, P~King$^{6,7}$, O~Lebeda$^8$, P~Lechner$^9$, V~S~Pantuev$^{10}$, D~S~Parno$^{11}$, M~Pavan$^{5,12}$, S~Pozzi$^{5,12}$, D~C~Radford$^{13}$, M~Slezák$^2$, M~Steidl$^3$, P~Trigilio$^4$, K~Urban$^{1,2}$, D~Vénos$^8$, J~Wolf$^3$, S~Wüstling$^3$ and Y-R~Yen$^{11}$}

\address{$^1$ Technical University of Munich, Arcisstraße 21, 80333 München, Germany}
\address{$^2$ Max Planck Institute for Physics, Föhringer Ring 6, 80805 München, Germany}
\address{$^3$ Karlsruhe Institute of Technology, Hermann-von-Helmholtz-Platz 1, 76344 Eggenstein-Leopoldshafen, Germany}
\address{$^4$ XGLab srl, Bruker Nano Analytics, Via Conte Rosso 23, 20134 Milano, Italy}
\address{$^5$ INFN - Sezione di Milano - Bicocca, 20126  Milano, Italy}
\address{$^6$ Politecnico di Milano, Piazza Leonardo da Vinci 32, 20133 Milano, Italy}
\address{$^7$ INFN - Sezione di Milano, 20133, Milano, Italy}
\address{$^8$ Nuclear Physics Institute of the CAS, v.\ v.\ i., 250 68 \v{R}e\v{z}, Czech Republic}
\address{$^9$ Halbleiterlabor of the Max Planck Society, Otto-Hahn-Ring 6, 81739 München, Germany}
\address{$^{10}$ Institute for Nuclear Research of Russian Academy of Sciences, Prospekt 60-letiya Oktyabrya 7a, Moscow 117312, Russian Federation}
\address{$^{11}$ Department of Physics, Carnegie Mellon University, Pittsburgh, Pennsylvania 15213, USA}
\address{$^{12}$ Universit\`a di Milano - Bicocca, Dipartimento di Fisica, 20126 Milano, Italy}
\address{$^{13}$ Oak Ridge National Laboratory, 1 Bethel Valley Road, Oak Ridge, TN 37831, USA}

\ead{mertens@mpp.mpg.de}

\vspace{10pt}
\begin{indented}
\item[]November 2020
\end{indented}

\begin{abstract}
Sterile neutrinos are a minimal extension of the Standard Model of Particle Physics.
A promising model-independent way to search for sterile neutrinos is via high-precision $\upbeta$-spectroscopy.
The Karlsruhe Tritium Neutrino (KATRIN) experiment, equipped with a novel multi-pixel silicon drift detector focal plane array and read-out system, named the TRISTAN detector, has the potential to supersede the sensitivity of previous laboratory-based searches.
In this work we present the characterization of the first silicon drift detector prototypes with electrons and we investigate the impact of uncertainties of the detector's response to electrons on the final sterile neutrino sensitivity.
\end{abstract}

\vspace{2pc}
\noindent{\it Keywords}: Solid state detectors, Particle detectors, Neutrinos, Sterile Neutrinos, KATRIN


%
%

\section{Introduction}
\label{chap:intro}

Sterile neutrinos are a minimal extension of the Standard Model of Particle Physics (SM)~\cite{Abazajian.2012}.
A common candidate for sterile neutrinos are right-handed partners to the standard left-handed neutrinos.
The existence of such right-handed partners would provide a natural way to introduce neutrino mass to the SM.
A consequence of this SM extension is the existence of new neutrino mass eigenstates.
These new neutrino particles can have an arbitrary mass scale $m_\mathrm{s}$ and a small admixture of the active neutrino component, governed by the so-called mixing amplitude $\sin^2(\Theta)$. This mixing allows them to interact with matter via the weak interaction. Throughout this work, we call these new neutrino mass eigenstates ``sterile'' neutrinos $\nu_\mathrm{s}$.

Light (eV-scale) sterile neutrinos are being widely discussed in the context of short-baseline neutrino oscillation anomalies, such as the reactor antineutrino anomaly~\cite{Mention:2011rk,Giunti:2019aiy,Boser:2019rta}. Very heavy sterile neutrinos of at least the GeV-scale are typically introduced to generate both neutrino masses and the matter/anti-matter asymmetry of the universe~\cite{Canetti:2012vf,Mohapatra:1979ia}. Finally, keV-scale sterile neutrinos are viable candidates for dark matter~\cite{Adhikari.2017, Boyarsky:2018tvu}.

Depending on their production mechanism sterile neutrinos can act as cold, cool, or warm dark matter, which would impact structure formation of the universe in different ways. In particular the warm type of sterile neutrino dark matter could potentially mitigate tensions between observations and cold-dark-matter predictions on small scales~\cite{Lovell.2011}. Sterile neutrinos constitute so-called decaying dark matter.
An interesting decay mode, from the observational point of view, is the decay to an active neutrino and a mono-energetic x-ray photon.
X-ray telescopes, such as the recently launched eROSITA~\cite{Merloni.2012}, can thus search for the existence of sterile neutrino dark matter.
Current x-ray data limit the mixing amplitude of sterile neutrinos to about $\sin^2(\Theta) < \numrange{e-11}{e-7}$ in a mass range of \SI{2}{keV} to \SI{10}{keV}, respectively~\cite{Boyarsky.2008,Watson.2012}.\footnote{All masses are stated in natural units.}
Below \SI{2}{keV} mixing amplitudes above \num{e-6} are disfavoured as they would lead to an overproduction of sterile neutrino dark matter.
However, these limits are rather model-dependent~\cite{Boyarsky.2014}.
Laboratory searches reach exclusion limits of $\sin^2(\Theta) < \numrange{e-4}{e-2}$ in a mass range of \SI{1}{keV} to \SI{90}{keV}~\cite{Abdurashitov.2017,Holzschuh.1999,Holzschuh.2000}.

A promising approach to search for sterile neutrinos in a laboratory-based experiment, which would be independent of cosmological or astrophysical models, is based on nuclear $\upbeta$-decays~\cite{Shrock.1980}.
In $\upbeta^-$-decays, an electron anti-neutrino $\bar{\nu}_\mathrm{e}$ is emitted alongside the $\upbeta$-electron.
Thus, the electron's energy spectrum is given as a superposition of spectra corresponding to the neutrino mass eigenstates $m_i$.
A mass eigenstate $m_\mathrm{s}$ in the keV-range would correspond to a spectral branch with a much reduced endpoint of $E = E_0 - m_\mathrm{s}$, where $E_0$ is the spectrum endpoint.
Accordingly the sterile neutrinos would generate a kink-like signature at this energy and a broad distortion of the spectrum at lower energies.
The relative probability of this decay branch is governed by the mixing amplitude $\sin^2(\Theta)$.

The large-scale Karlsruhe Tritium Neutrino (KATRIN) experiment is currently the world-leading facility for precision spectroscopy of tritium.
Its main objective is the direct measurement of the effective electron anti-neutrino mass.
In 2019, the collaboration set an improved upper limit of $m(\bar{\nu}_\mathrm{e}) < \SI{1.1}{\eV}$ at a confidence level (C.L.) of \SI{90}{\%}~\cite{Aker.2019}.
The activity of its tritium source is ultra-high (\num{e11} decays per second) and stable (\SI{0.1}{\%} per hour), allowing to extend KATRIN's physics program to search for keV-scale sterile neutrinos~\cite{Mertens.2015}.
To enable a sterile-neutrino measurement, an upgrade of the KATRIN apparatus with a novel multi-pixel focal plane detector array and read-out system is necessary.
The three main challenges this system faces are 1) to handle high rates at the level of up to \SI{e8}{cps}, 2) to provide an energy resolution of about \SI{300}{\eV} full width at half maximum (FWHM) at \SI{20}{\keV} for electrons, and 3) to control the energy linearity at the ppm-level.

In the framework of the Tritium Investigations on Sterile-to-Active Neutrino Mixing (TRISTAN) project, first prototype detectors based on the Silicon Drift Detector (SDD) technology have been developed and tested~\cite{Mertens.2019}.
SDDs are ideally suited for high-rate and high-energy-resolution applications.
Thanks to their small read-out anode, even large pixel sizes of up to several millimeters keep a small capacitance at the level of few hundreds of \si{\femto\farad}~\cite{Lechner.2001}, which leads to a low serial noise.
This in turn allows for short energy filter shaping times and is advantageous for measurements at high rates.
Typically, SDDs are used for x-ray measurements.
In the case of TRISTAN these detectors will be applied for high-precision electron spectroscopy.
In contrast to the x-ray application, the effects of energy loss in an insensitive region at the entrance window and backscattering from the detector's surface play a major role for the detection of electrons. 

In this work we present a detailed characterization of SDDs with mono-energetic electrons.
A scanning electron microscope (SEM) and a \rbkr radioactive source were used as calibration sources (see sec.~\ref{chap:measurement}).
Based on the obtained spectra an empirical model describing the detector's response to electrons was developed (see sec.~\ref{chap:response}).
A special focus of this work is put on the estimation of the entrance-window thickness (see sec.~\ref{chap:thickness}). Finally, the impact of uncertainties in the electron-response model on the final sensitivity to sterile neutrinos is presented in sec.~\ref{chap:impact}.

\section{Characterization with electrons}
\label{chap:measurement}
A novelty of the TRISTAN detector system is the application of SDDs to high-precision electron spectroscopy.
The energy deposition profile is different for massive charged particles and photons.
This fact has several consequences:
First, electrons deposit a significant fraction of their energy close to the entrance-window surface, where the electric fields are too weak to transport the charge carriers to the read-out anode.
Thus the energy of the electron is not fully detected but partly lost.
Second, low-energy electrons have a probability of around \SI{20}{\%} or larger to scatter back from the silicon detector surface, again leading to a partial energy measurement~\cite{Darlington.1972}.
Finally, due to the interactions close to the surface, characteristic \SI{1.74}{\keV} Si x-rays created by the electron can escape the detector volume, again reducing the detected energy of the electron.
All these effects lead to a characteristic shape of the energy spectrum for electrons.

To characterize the response of SDDs to electrons, a 7-pixel TRISTAN prototype and two mono-energetic electron sources were used.
In the following we describe the detector system and the two calibration sources, and present the obtained spectra.
All measurements described in this work were performed at room temperature.

\subsection{Prototype detector}
\label{sec:Det}

\begin{figure}
    \centering
    \subfloat[Read-out side.]{
    	\includegraphics[width=0.45\textwidth]{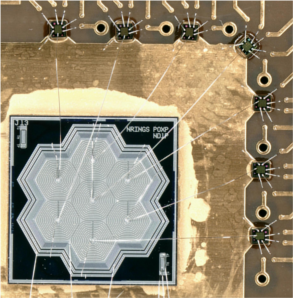}
    	\label{fig:Detector_front}
    }
    \subfloat[Entrance-window side.]{
    	\includegraphics[width=0.45\textwidth]{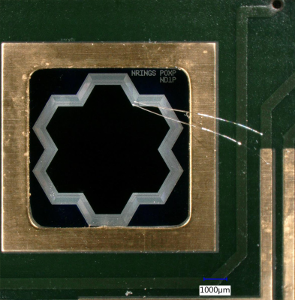}
   		\label{fig:Detector_back}
   	}
    \caption{Photographs of the TRISTAN detector chip.
    \protect\subref{fig:Detector_front} The read-out side of the detector chip. Each pixel anode is surrounded by twelve drift rings and bonded to a CUBE preamplifier ASIC. The pixel size is \SI{2}{\mm} and the edge length of the silicon chip is \SI{8}{\mm}.
    \protect\subref{fig:Detector_back} The entrance-window side of the detector chip shows no structuring into individual pixels. Depletion voltage and guard-ring voltage are supplied via wire bonds.}
    \label{fig:Detector}
\end{figure}

Several 7-pixel SDD arrays have been manufactured at the Semiconductor Laboratory of the Max Planck Society (HLL)~\cite{Lechner.2001}.
The chips are produced from monolithic silicon wafers with a thickness of \SI{450}{\micro\meter}.
They feature a thin entrance window, terminated by a \SI{10}{\nano\meter} thick SiO$_2$ layer.
The hexagonal pixel shape enables an arrangement without dead area. For this work SDD chips with \SI{2}{\milli\meter} pixel diameter, shown in fig.~\ref{fig:Detector}, were used.
Each of these pixels features twelve drift rings and a small anode capacitance of approximately \SI{100}{\femto\farad}.

Each anode is wire bonded to a charge-sensitive preamplifier (CSA) application-specific integrated circuit (ASIC), situated in close vicinity to the chip.
The ``CUBE'' ASIC has been developed by Politecnico di Milano and XGLab for applications in high-count rate and low-noise spectroscopy~\cite{Bombelli.2011, Lechner.2001}.
Its field effect transistor (FET) is based on complementary metal oxide semiconductor (CMOS) technology and operates in pulsed-reset mode.

The ``DANTE'' digital pulse processor (DPP) is used as a back-end electronics.
DANTE is provided by XGLab and features a waveform-digitizing analog-to-digital converter (ADC) with a sampling frequency of \SI{125}{\mega\hertz} and \SI{16}{bit} resolution.
Two trapezoidal filters are applied for event triggering and energy reconstruction, optimizing the system for high-count-rate applications~\cite{Jordanov.1994}.
Characterization measurements of the TRISTAN SDD with a $^{55}$Fe x-ray source demonstrated an energy resolution of \SI{139}{\eV} (FWHM) at \SI{5.9}{\keV} with trapezoidal filter peaking times of about \SI{1}{\micro\second}~\cite{Mertens.2019} and a detector temperature of \SI{-30}{\celsius}.

\subsection{Calibration sources}
\label{sec:Sources}

\paragraph{\rbkr source}

The krypton calibration source consists of a mono-layer of \rb, evaporated onto a highly oriented pyrolytic graphite (HOPG) carrier substrate~\cite{Venos.2010}. \rb decays with a half-life of 86.2 days via electron capture to \kr{}.
In this decay, predominantly Kr-K\textsubscript{\textalpha} and Kr-K\textsubscript{\textbeta} x-rays are emitted. The occurrence of both photon and electron lines in one spectrum enables an in-situ calibration and characterization at the same time. The isomeric state \kr{} is the second excited state of krypton and has an energy of about \SI{41.6}{\keV}~\cite{McCutchan.2015}. 
It deexcites via a cascade of two \textgamma-decays with \SI{32.2}{\keV} and \SI{9.4}{\keV}, respectively. 
In both transitions, conversion electrons occur with energies
\begin{equation}
    E_\mathrm{ce} = E_\upgamma + E_\upgamma\mathrm{(recoil)} - E_\mathrm{ce}\mathrm{(bind)} - E_\mathrm{ce}\mathrm{(recoil)}~,
\end{equation}
defined by the energy of the $\upgamma$-transition $E_\upgamma$, the binding energy of an electronic shell of the krypton atom $E_\mathrm{ce}\mathrm{(bind)}$, and the recoil energies of the atom after the emission of the $\upgamma$-ray $E_\upgamma\mathrm{(recoil)}$ and the conversion electron $E_\mathrm{ce}\mathrm{(recoil)}$.
The emission of a conversion electron from the K-shell ($E_\mathrm{ce}\mathrm{(bind)}=\SI{14.3}{\keV}$) is energetically forbidden in the second deexcitation.
Photon and electron peak energies of specific interest for this work are listed in tab.~\ref{tab:Krlines}.
All other lines are situated in a low-energy continuum and thus unsuited for the investigation.
\begin{table}
    \caption{Listed are \protect\subref{tab:KrlinesPh} photon and \protect\subref{tab:KrlinesEl} electron lines of \rbkr~\cite{Venos.2018} used for the analysis in this work. Some lines are not resolvable given the energy resolution of the detector and appear as a single peak in the spectrum.}
	\label{tab:Krlines}
	\subfloat[][Photons]{
    \begin{minipage}[t]{.45\linewidth}
    	\centering
    	\sisetup{table-number-alignment=center, 
    			 table-figures-decimal=3,
    			 add-decimal-zero,
    		     table-format = -5.2(2)
    		     }
    		\begin{tabular}{clS}
    			\toprule
    			Peak                            & {Line}                            & {Energy (\si{\eV})}   \\
    			\midrule
    			\midrule
    			\textgamma{}-9.4                & \textgamma{}-9.4                  & 9405.7\pm 0.6         \\
    			\midrule
    			\multirow{2}{*}{\kalpha{}}   & K\textsubscript{\textalpha{}2}    & 12595.4\pm 0.1        \\
    											& K\textsubscript{\textalpha{}1}    & 12648.0\pm 0.1        \\
    			\midrule
    			\multirow{3}{*}{\kbeta{}}    & K\textsubscript{\textbeta{}3}     & 14105.0\pm 0.1        \\
    										    & K\textsubscript{\textbeta{}1}     & 14112.8\pm 0.1        \\
    										    & K\textsubscript{\textbeta{}2}     & 14315.0\pm 2.4        \\
    			\bottomrule
    			&&\\
    			&&\\
    			&&\\
    		\end{tabular}
    	\label{tab:KrlinesPh}%
	\end{minipage}}%
	\subfloat[][Electrons]{\begin{minipage}[t]{.45\linewidth}
	    \centering
    	\sisetup{table-number-alignment=center, 
    		     table-figures-decimal=3,
    		 	 add-decimal-zero,
    		     table-format = -5.1(1)
    		     }
    		\begin{tabular}{clS}
    			\toprule
    			Peak                                    & {Line}            & {Energy (\si{\eV})}   \\
    			\midrule
    			\midrule
    			\krline{K}{}{32}                        & \krline{K}{}{32}  & 17824.2\pm 0.5        \\
    			\midrule
    			\multirow{3}[0]{*}{\krline{L}{}{32}}    & \krline{L}{1}{32} & 30226.8\pm 0.9        \\
    			                                        & \krline{L}{2}{32} & 30419.5\pm 0.5        \\
    		                                            & \krline{L}{3}{32} & 30472.2\pm 0.5        \\
    			\midrule
    			\multirow{5}[0]{*}{\krline{M,N}{}{32}}  & \krline{M}{1}{32} & 31858.7\pm 0.6        \\
                                            			& \krline{M}{2}{32} & 31929.3\pm 0.5        \\
                                            			& \krline{M}{3}{32} & 31936.9\pm 0.5        \\
                                            			& \krline{N}{2}{32} & 32136.7\pm 0.5        \\
                                            			& \krline{N}{3}{32} & 32137.4\pm 0.5        \\
    			\bottomrule
    		\end{tabular}
	    \label{tab:KrlinesEl}%
	\end{minipage}}%
\end{table}%

\begin{figure}
    \centering
	\includegraphics[width=0.7\linewidth]{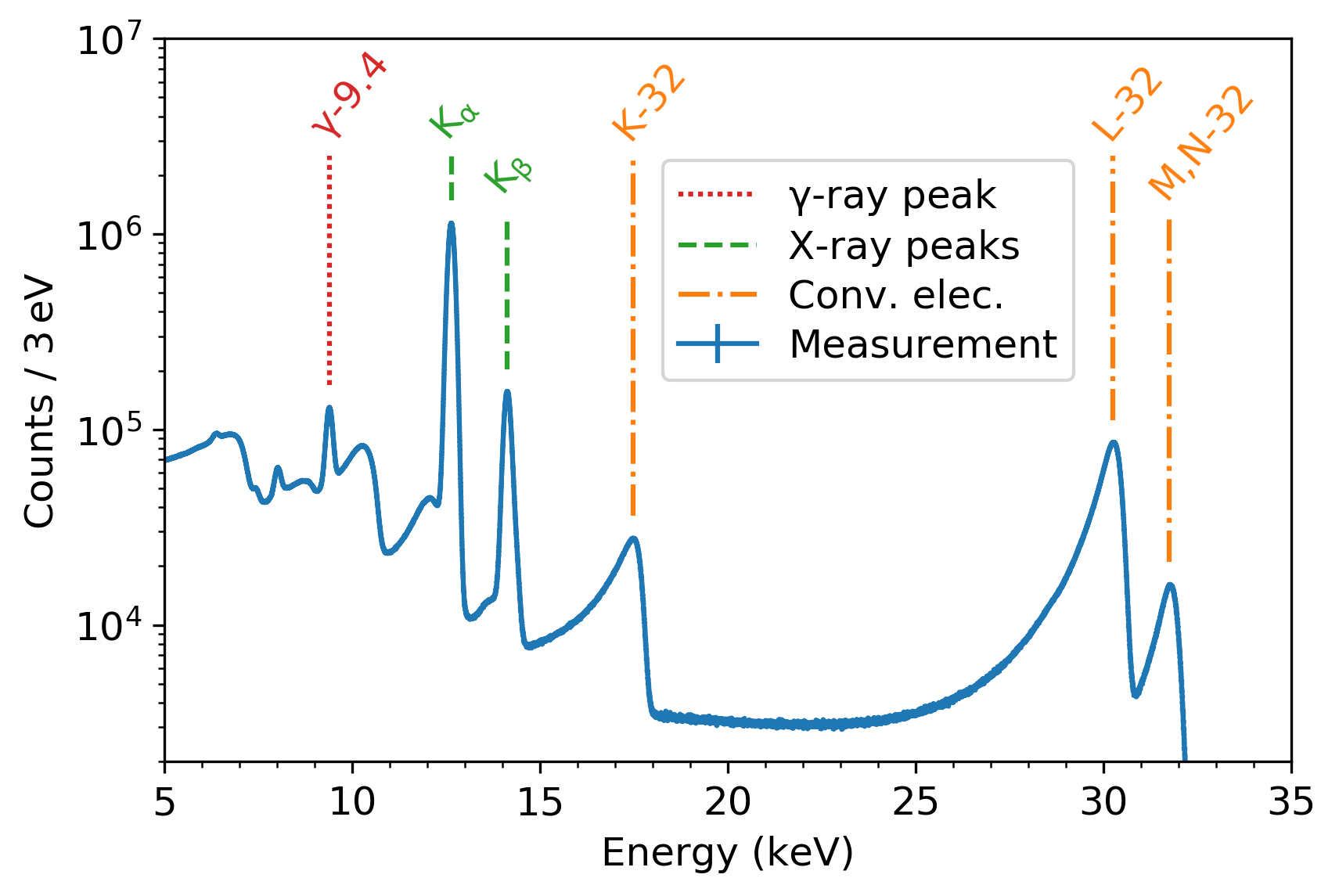}
    \caption{Spectrum of a \rbkr source measured with a TRISTAN prototype detector. The peaks of specific interest for this work are labeled.}
    \label{fig:krspectrum}
\end{figure}

During the measurement, the source was positioned about \SI{1}{\cm} underneath the detector entrance window, such that the entire pixel array was illuminated homogeneously.
A typical spectrum of \rbkr measured with the TRISTAN SDD detector is shown in fig.~\ref{fig:krspectrum}. 
While the photon peaks are symmetric, peaks from conversion electrons show a pronounced low-energy tail.
Furthermore, the electron peak position $\bar{E}_i$ of a peak $i$ is shifted towards lower energies compared to the respective theoretical value $\bar{E}^\mathrm{th}_i$:
\begin{equation}
    \bar{E}_i = \bar{E}^\mathrm{th}_i - \Delta_i^\mathrm{ew} - \delta_\mathrm{sc} - \delta_{\Phi}~.
    \label{eq:PeakPos}
\end{equation}
This shift is due to energy losses in the entrance window $\Delta_i^\mathrm{ew}$ of the detector, potential energy losses in the source $\delta_\mathrm{sc}$, and the potential difference between source and entrance window $\delta_\Phi$~\cite{Lebert.2020}.

\paragraph{Scanning electron microscope}

Scanning electron microscopes (SEM) are generally used to visualize small structures, which cannot be resolved optically. In this work the SEM was used to generate mono-energetic electrons as calibration source for the TRISTAN SDD detector. 

\begin{figure}
    \centering
    \includegraphics[width=0.7\textwidth]{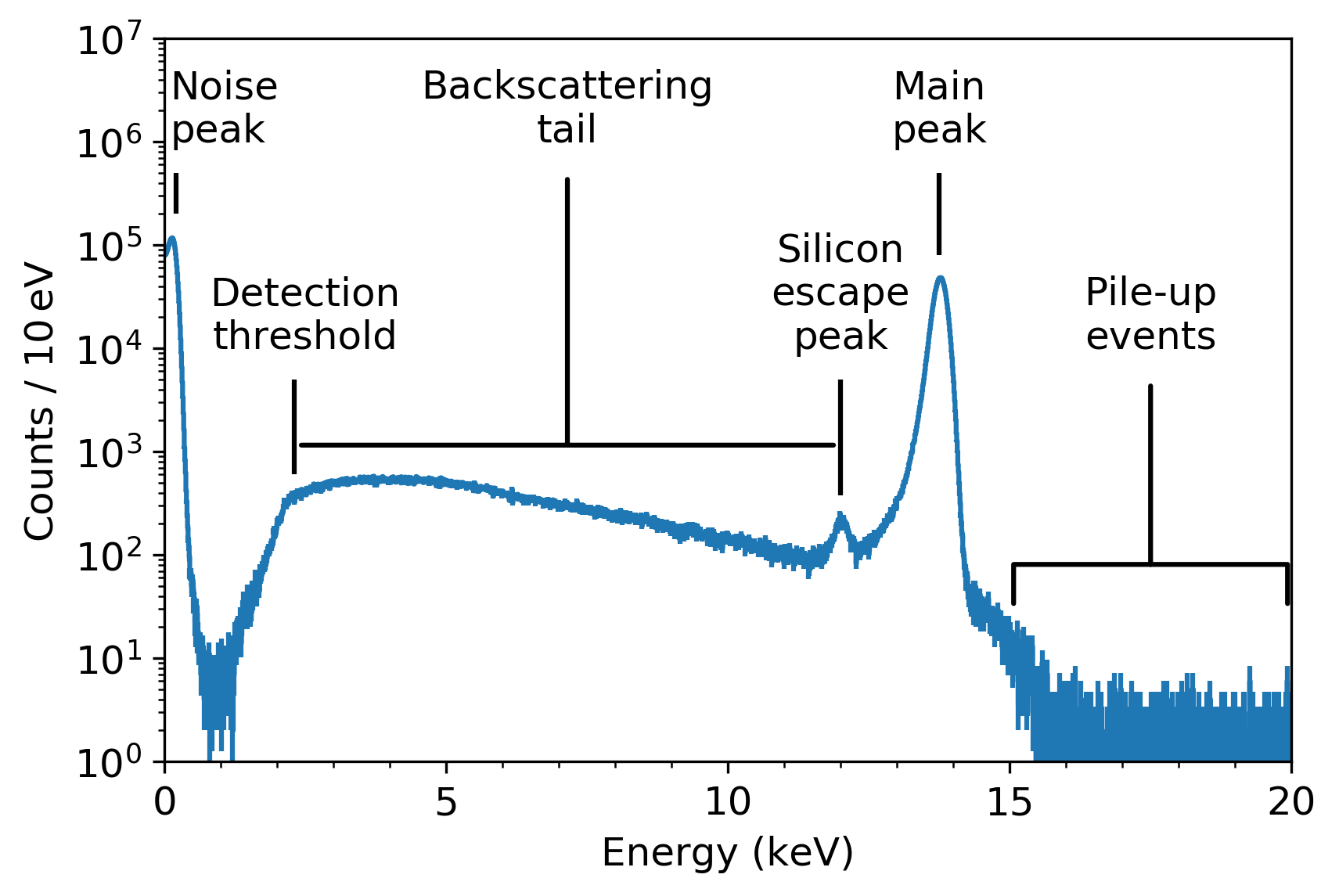}
    \caption{Spectrum of a \SI{14}{\keV} mono-energetic electron beam from an electron microscope measured with a TRISTAN prototype detector.}
    \label{fig:MonoSpec}
\end{figure}

An image area is repeatedly scanned with an electron beam of about \SI{10}{\nm} diameter.\footnote{JEOL JSM-IT300}
The electrons are accelerated to kinetic energies of up to \SI{30}{\keV} and are focused by a fast-changing magnetic field onto the sample inside a vacuum chamber.
The beam intensity is determined by the temperature of a heated tungsten spiral, from which the electrons are emitted.
The TRISTAN detector was positioned on the sample holder and electrically connected to the DAQ system via a feedthrough in a flange of the chamber.
First investigations with a single channel detector have shown a good applicability of this method~\cite{Gugiatti.2020}.
During the measurement of the 7-pixel detector, the beam rapidly scanned over the entire array, averaging the response over all positions of beam incidence.
A typically recorded electron energy spectrum is displayed in fig.~\ref{fig:MonoSpec}.

\section{Empirical model of the detector response}
\label{chap:response}
In order to describe the SDD response to electrons, an empirical analytical model was developed, where each physical effect is modelled by a separate term. In the following the main features of the spectrum are explained and their corresponding analytical expression is given.

\subsection{Spectral components}
\label{sec:effects}

\paragraph{Main peak}

The majority of electrons deposit almost their entire initial energy in the sensitive volume of the detector, leading to a main peak in the energy spectrum. Its general shape is well approximated by a Gaussian function
\begin{equation}
I_\mathrm{G}(E)  = A_G \cdot \exp\left(- \frac{\left(E - \mu\right)^2}{2 \sigma^2}\right),
\end{equation}
where $A_G$ is the amplitude, $\mu$ is the mean and $\sigma$ is the standard deviation. 

\paragraph{Low-energy tail}

The entrance-window surface is covered with a \SI{10}{\nano\meter} thick silicon oxide (SiO$_2$) layer.
Charge deposited in this layer cannot be detected.
Moreover, the electric fields in the silicon volume close to this layer are too weak to efficiently transport the charge carriers to the read-out contact.
These undetected energy depositions lead to an asymmetry of the main peak at its low-energy shoulder, which is modelled with the following function: 
\begin{equation}
    I_\mathrm{D}(E) = A_{D} \cdot \exp\left(\frac{E-\mu}{\beta}\right)\left(1 - \text{erf}\left[\frac{E - \mu}{\sqrt{2\sigma^2}} + \frac{\sigma}{\sqrt{2} \beta}\right]\right)~.
\end{equation}
It is composed of a ``washed-out'' step function expressed by the error function and an exponential tail towards lower energies.
Amplitude and slope of the function are given by $A_D$ and $\beta$, respectively.

\paragraph{Silicon escape peak}

Incident radiation leads to the ionization of silicon atoms in the detector material, most often on the K-shell.
In the subsequent $\mathrm{K}_\upalpha$ de-excitation, an x-ray with \mbox{$\Delta E_\mathrm{esc}=\SI{1.74}{\keV}$} is emitted.
If this photon leaves the detector, the energy $\Delta E_\mathrm{esc}$ remains undetected.
Hence, the silicon escape peak is modeled as a scaled projection of the main peak with amplitude $A_{\mathrm{esc}}$, shifted towards lower energies by $\Delta E_\mathrm{esc}$:
\begin{equation}
    I_{\mathrm{esc}}(E) = A_{\mathrm{esc}} \cdot  \exp{\left(-\frac{\left(E - \left[\mu - \Delta E_{\mathrm{esc}}\right]\right)^2}{2\sigma^2}\right)}~.
\end{equation}

\paragraph{Backscattering tail}

A fraction of primary and secondary electrons scatter back from the detector surface or escape after incomplete energy deposition.
The backscattering probability for electrons with energies of around \SI{20}{\keV} is about \SI{20}{\%} at perpendicular electron incidence and increases with the incident angle~\cite{Darlington.1972}.
This effect leads to a backscattering tail, dominating the spectral shape between silicon escape peak and detection threshold.
The probability for an electron to be backscattered is the largest at the detector surface and decreases with increasing penetration into the detector material.
Consequently, the backscattering tail rises towards lower energies.
We describe this tail with a multiplication of two power functions given by
\begin{equation}
    I_\mathrm{B}(E) = A_B \cdot \left(\frac{E}{\mu - a}\right)^b \cdot \left(1 - \frac{E}{\mu}\right)^c~.
\end{equation}
The first term with exponent $b$ describes the low-energy region just above the threshold energy $a$, whereas the second term with exponent $c$ describes the higher end of the backscattering tail.
Parameter $A_B$ is the overall amplitude of the function.

\subsection{Fit to the data}
\label{sec:SpecFit}

\begin{figure}
    \centering
    \subfloat[Fit to a krypton \krline{K}{}{32} peak.]{
    	\includegraphics[width=0.45\textwidth]{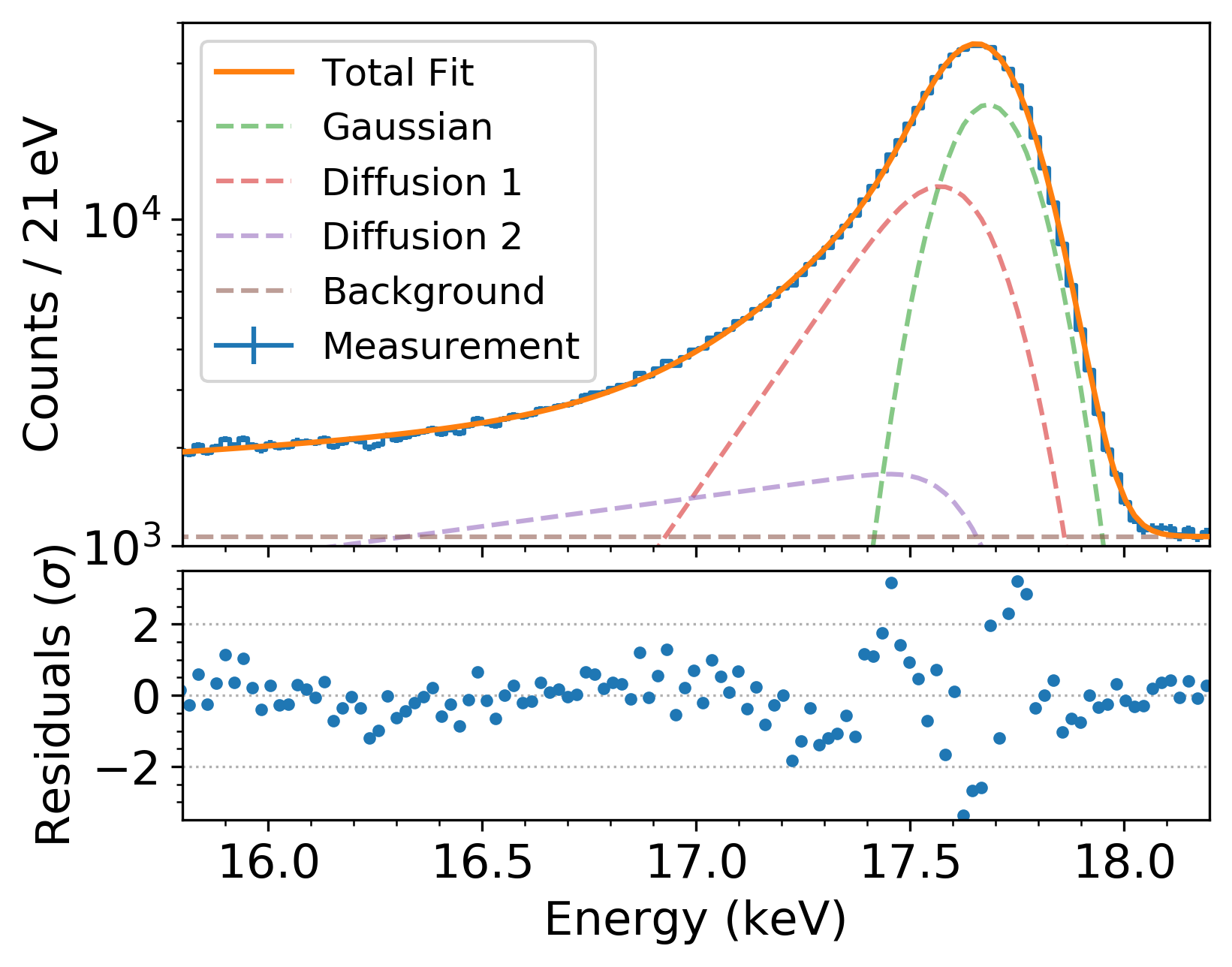}
    	\label{fig:Modelfit_Kr}
    }
    \subfloat[Fit to an electron microscope spectrum.]{
    	\includegraphics[width=0.45\textwidth]{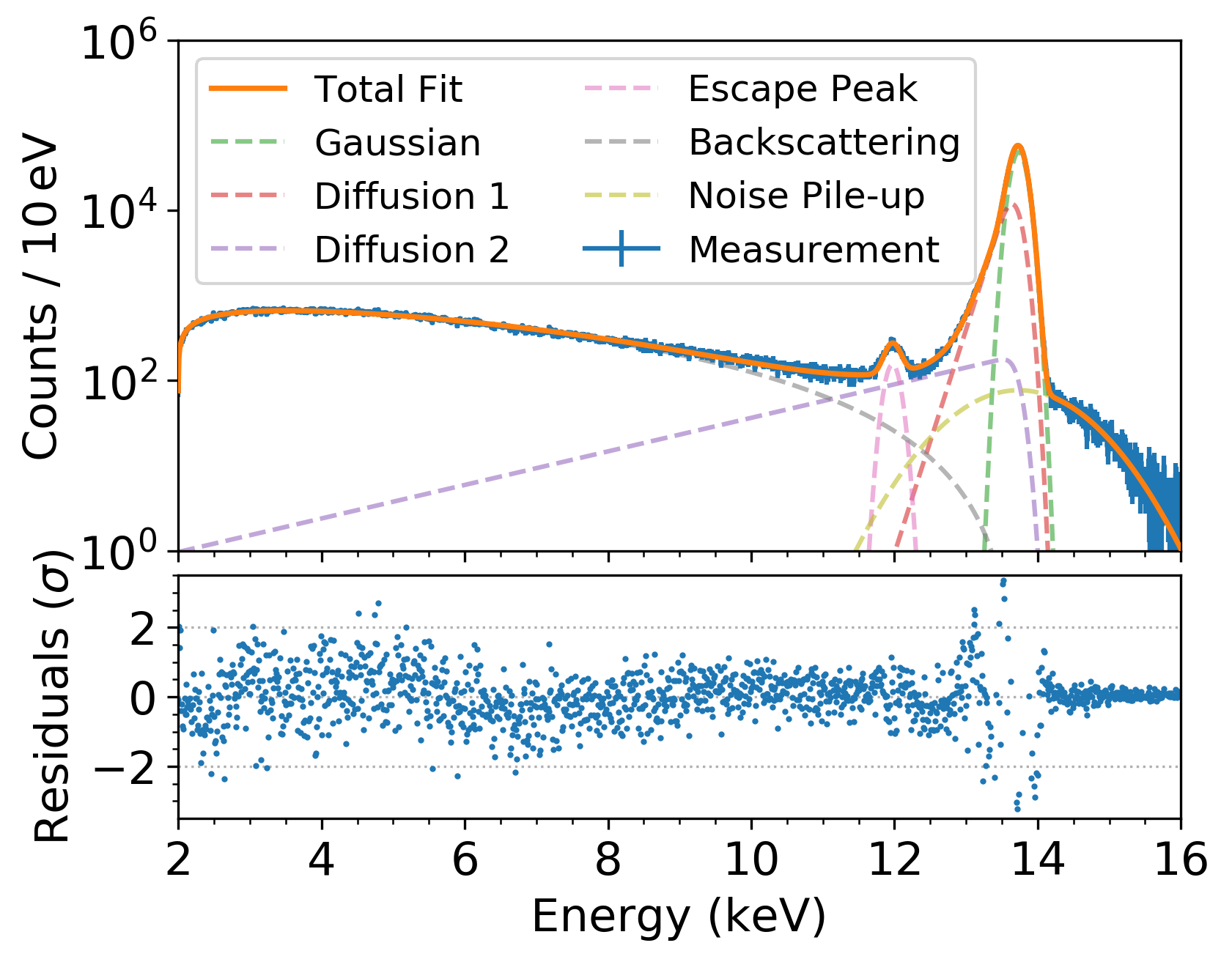}
   		\label{fig:Modelfit_SEM}
   	}
    \caption{Fits of the electron response model to measured spectra.
    \protect\subref{fig:Modelfit_Kr} The krypton K-32 line at around \SI{17.8}{\keV} is marked in fig.~\ref{fig:krspectrum}. All subfunctions of the model add up to a good fit ($\upchi^2/ \mathrm{dof}=122/ 90$) to the measured data.
    \protect\subref{fig:Modelfit_SEM} The energy of the SEM was set to \SI{14}{\keV}. Silicon escape peak and noise pile-up are approximated with two additional functions. The fit yields $\upchi^2/ \mathrm{dof}= 1958/ 1500$.}
    \label{fig:Modelfit}
\end{figure}

Each mono-energetic electron peak measured with \kr{} or at the electron microscope is fit with the sum of the terms described above.
In the case of the \krline{K}{}{32} and \krline{M}{}{32} peaks, the terms $I_B(E)$ and $I_\mathrm{esc}(E)$ are not considered, as the fit is only performed in a region close to each conversion electron peak because various peaks are overlapping. 
An example of the fit to the \krline{K}{}{32} peak is shown in fig.~\ref{fig:Modelfit_Kr}.
The fit to the obtained spectrum at the electron microscope measurement is displayed in fig.~\ref{fig:Modelfit_SEM}. 

A very good agreement of the empirical model with the data is found. The dependence of the model parameters on both energy and angle can be investigated by measuring the response at various energies and incident angles.
For the analysis of the continuous tritium $\upbeta$-decay spectrum, it is conceivable to use this empirical description of the detector response to obtain a model of the measured tritium spectrum.
To this end, the mono-energetic spectra are combined into a response matrix, which is then multiplied by the theoretical $\upbeta$-decay spectrum.
A parameterization of the response is particularly advantageous as it allows one to easily include systematic uncertainties in the data analysis.
The feasibility of this approach has been demonstrated in~\cite{Brunst.2019}.

\section{Estimation of the entrance-window thickness}
\label{chap:thickness}
In this section we focus on a detailed investigation of the entrance-window thickness.
As described above, charge carriers created in a volume close the detector surface are only partially collected.
Approximately the first \SI{10}{\nm} of the detector are fully insensitive, due to a SiO$_2$ layer.
In the silicon bulk the charge-collection efficiency increases steeply.
In this work, we assume a sharp transition between the dead and active detector areas for simplicity, which we refer to as a ``dead-layer model''.

In the following we describe the measurement strategy and we present the resulting estimation of the dead-layer thickness, which is obtained by comparing the measurement results to Monte Carlo (MC) simulations.

\subsection{Tilted beam method}
\label{sec:KrTilt}

\begin{figure}
  \centering
  \includegraphics[width=0.7\linewidth, page=4]{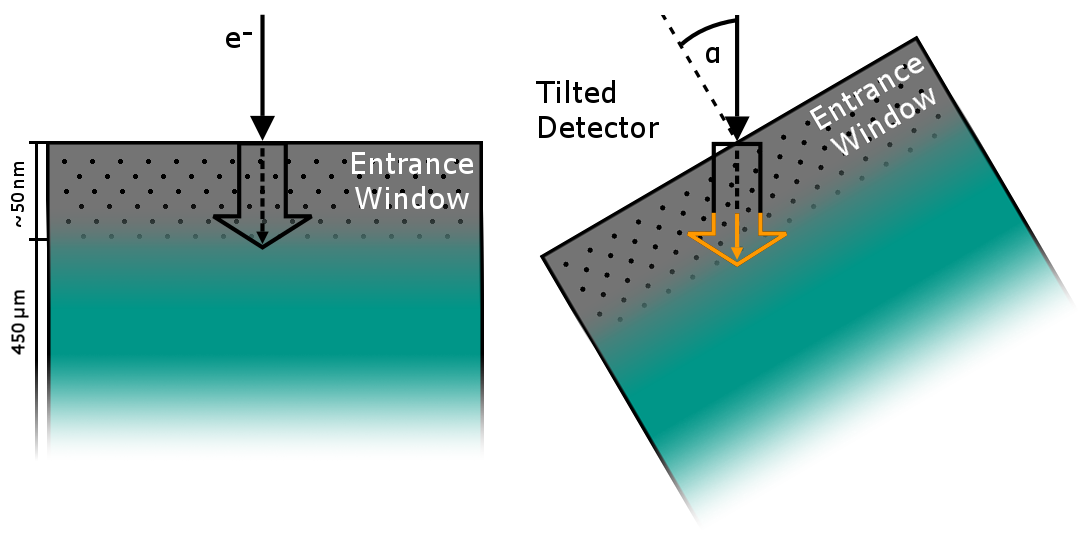}
  \caption[Scheme of the tilted beam method]{Scheme of the tilted beam method.
  The distance that electrons travel through the entrance window (grey region) before reaching the sensitive detector volume (green region) is effectively increased (orange arrow) by tilting the detector by an angle $\alpha$.}
  \label{fig:Tilt_Motivation}
\end{figure}

To eliminate possible influences of the bias voltage $\delta_{\Phi}$ and source effects $\delta_\mathrm{sc}$ on the peak position shift (see equ.~\ref{eq:PeakPos}), the tilted beam method is applied~\cite{Johansen.1990}.
By tilting the detector relative to the electron source by an angle $\alpha$, the effective dead-layer thickness for incoming electrons increases.
This concept is illustrated in fig.~\ref{fig:Tilt_Motivation}.
By comparing measurements with and without tilt angle, the influence of the entrance window $\Delta_i^\mathrm{ew}$ is isolated as a relative shift of the main energy peaks: 
\begin{equation}
    \Delta E = E(\alpha) - E(\ang{0})~.
\end{equation}
Measurements were performed with the scanning electron microscope and the \hbox{\rbkr} source with perpendicular electron incidence ($\alpha=\ang{0}$) and tilted detector by $\alpha=\ang{60}$.
The measured spectra are fit with the empirical response model described in sec.~\ref{sec:effects}.
The resulting peak position differences $\Delta E$ are calculated and illustrated in fig.~\ref{fig:rsims_data}. For a 14~keV electron $\Delta E = 50$~eV, and, as expected, the energy loss decreases to about $\Delta E = 30$~eV at incident energies of about 30 keV.

\subsection{Interpretation as entrance-window thickness}
\label{sec:Interpretation}
In this study we describe the region of incomplete charge collection at the entrance window with a single parameter, the dead-layer thickness $d_{\mathrm{DL}}$.
To relate the observed shift in energy $\Delta E$ to a dead-layer thickness, MC simulations of electrons are performed with two incident angles of $\alpha=\ang{0}$ and $\alpha=\ang{60}$ and for several incident energies $E_i$.
For each case the simulation is performed with eleven different dead-layer thicknesses in a range of \SIrange{40}{60}{\nm} in steps of \SI{2}{\nm}.
The \SI{10}{\nano\meter} thick Si$\mathrm{O}_2$ layer is included in this dead layer model.
The simulations are performed with the KESS software, which was developed by the KATRIN collaboration specifically to describe scattering of low-energy electrons in silicon~\cite{Chaoui.2009,Chaoui.2010,Chaoui.2010_auger,Chaoui.2010_edep,Renschler.2011}. 

\begin{figure}
    \centering
    \subfloat[Energy dependent peak position shift.]{
    	\includegraphics[width=0.45\textwidth]{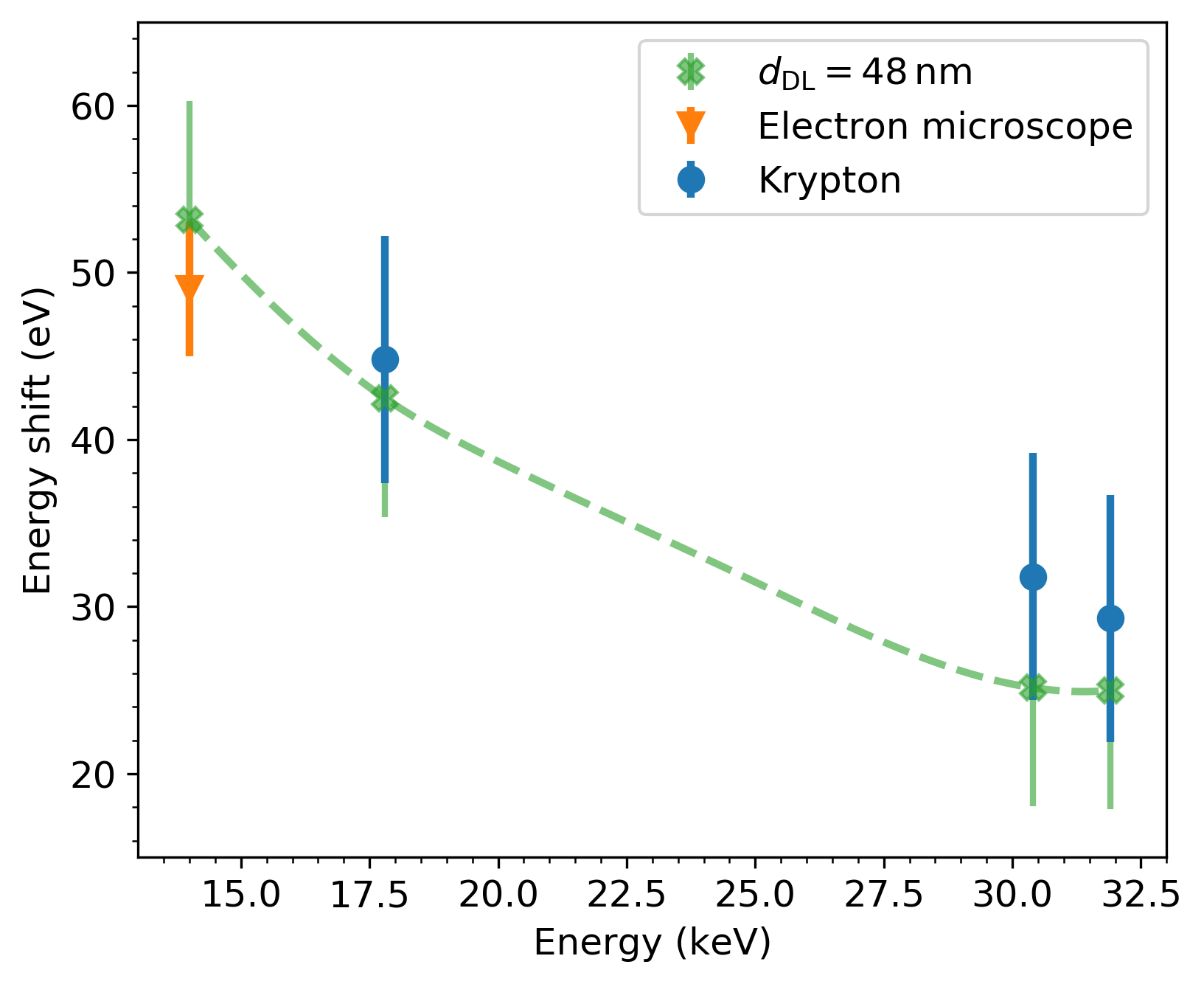}
    	\label{fig:rsims_data}
    }
    \subfloat[Simulated and measured spectrum.]{
    	\includegraphics[width=0.45\textwidth]{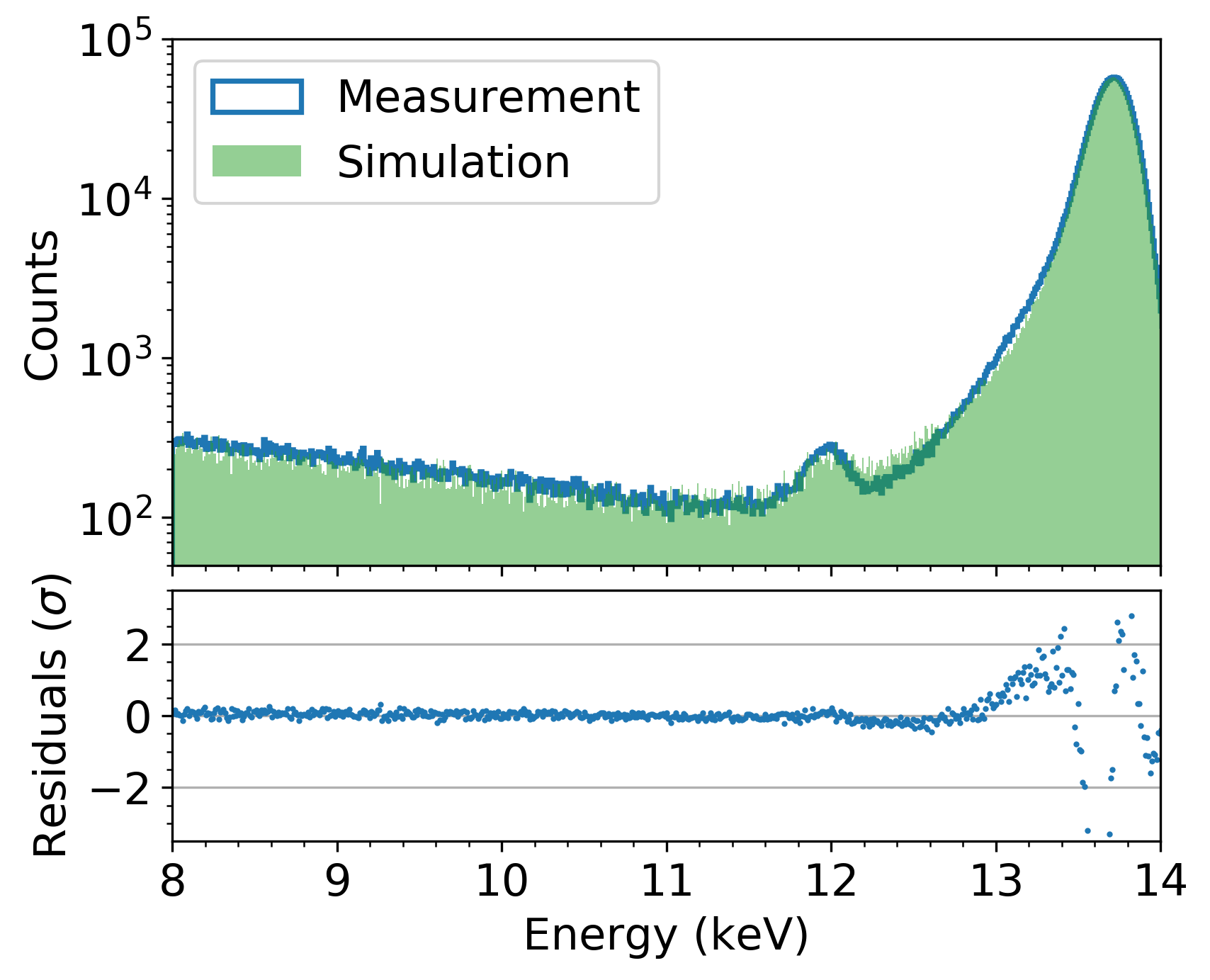}
   		\label{fig:S0-Simulation_vs_Data_14keV_S0-2}
   	}
    \caption{Comparison of the measurements with MC simulations at perpendicular electron incidence using KESS.
    \protect\subref{fig:rsims_data} Energy shifts extracted from measurements with an electron microscope and a \rbkr source. The values are compared to a simulation based on a dead-layer model with $d_\mathrm{DL}=\SI{48}{\nano\meter}$.
    \protect\subref{fig:S0-Simulation_vs_Data_14keV_S0-2} Spectrum of mono-energetic electrons from a measurement with an electron microscope and the corresponding simulation.}
    \label{fig:DLextraction}
\end{figure}

Fig.~\ref{fig:S0-Simulation_vs_Data_14keV_S0-2} shows a good agreement of the measured and a MC simulated spectrum.
A minimization of the squared data-to-simulation residuals for each measured energy yields the best dead-layer thickness and uncertainty of
\begin{equation}
    d_\mathrm{DL} = \SI{49 +- 3}{\nm}~.
    \label{equ:dl_thick_meas}
\end{equation}
Using this dead-layer thickness, an overlay of the simulated and measured energy shifts is illustrated in fig.~\ref{fig:rsims_data}.

\section{Impact of the detector response on the sterile neutrino sensitivity}
\label{chap:impact}
The final aim of the TRISTAN project is to reach a sensitivity to a spectral distortion at the ppm-level.
This requires two key ingredients: 1) an excellent energy resolution and 2) a precise understanding of the measured  spectral shape.
In the following, we discuss the impact of the entrance-window thickness on the energy resolution, and secondly the impact of an imprecise knowledge of the detector response to electrons on the sterile-neutrino sensitivity.

\subsection{Impact on energy resolution}
\label{sec:EnergyRes}
To reach the targeted sterile-neutrino sensitivity an energy resolution of approximately \SI{300}{\eV} at \SI{20}{\keV} is required~\cite{Mertens.2015.Wavelet}.
An excellent energy resolution is needed to avoid washing out the characteristic kink-like signature of a sterile neutrino.
The ability to detect or rule out this local signature makes the search for sterile neutrinos robust against large classes of systematic uncertainties.

\begin{figure}
    \centering
	\includegraphics[width=0.7\linewidth]{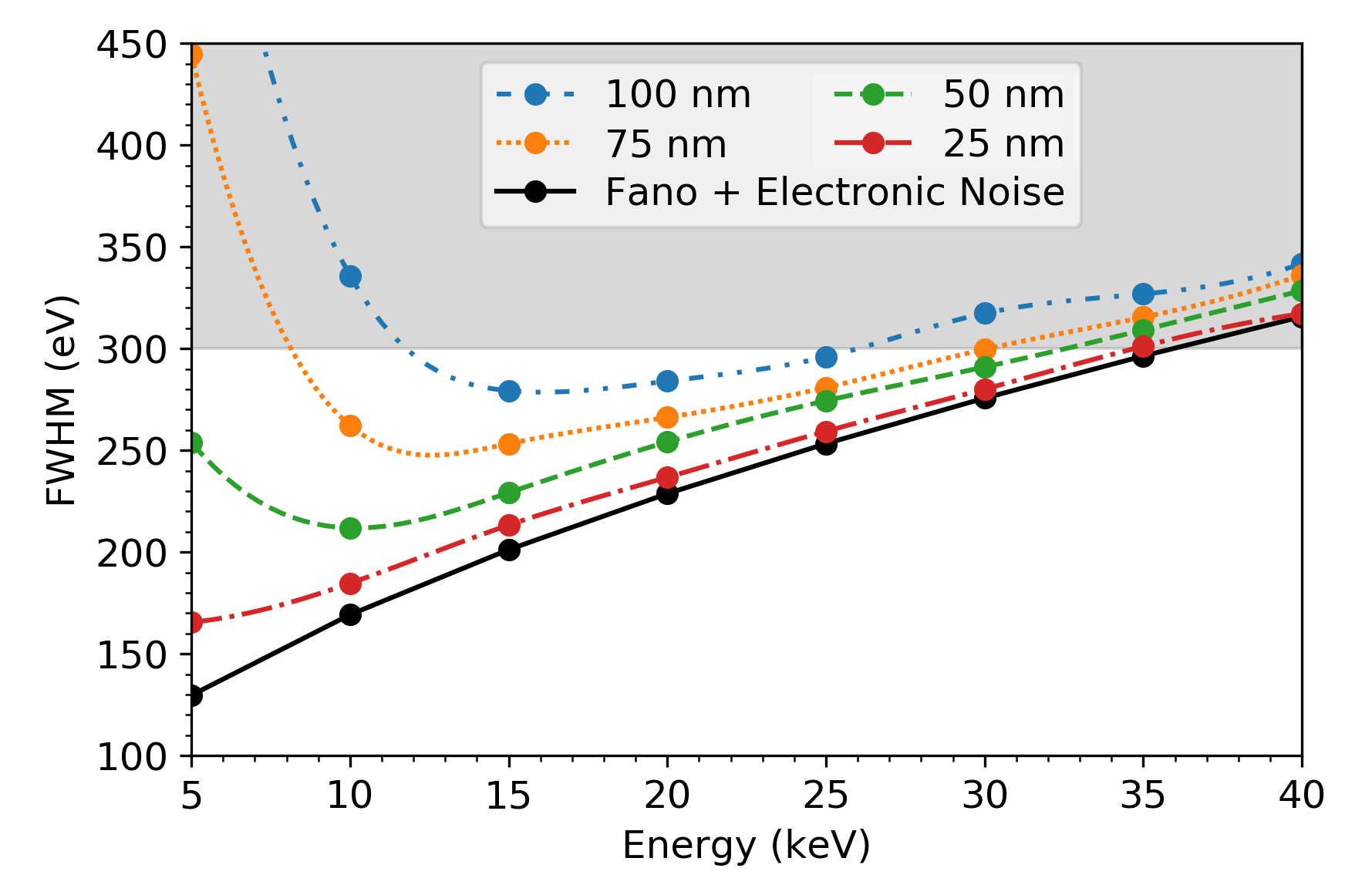}
    \caption{Simulation of the additional broadening of the energy resolution in FWHM, for different dead-layer thicknesses. An equivalent noise charge of \num{9} electrons is considered~\cite{Mertens.2019}. To reach the targeted sensitivity a FWHM of less than \SI{300}{\eV} is required.}
    \label{fig:Eres}
\end{figure}

Energy loss in the dead layer is a statistical process and thus different for each incident electron.
This variation leads to a smearing of the spectrum and can be interpreted as a worsening of the energy resolution.
Fig.~\ref{fig:Eres} illustrates the resulting additional contribution to the energy resolution for different dead-layer thicknesses. This contribution is especially large at low energies, where the energy loss in the dead layer is the largest.
With the measured effective dead-layer thickness (see equ.~\ref{equ:dl_thick_meas}) the requirement on energy resolution is met.

\subsection{Impact on final sensitivity}
\label{sec:DLBS}
To reach ppm-level sensitivity to sterile neutrinos, systematic effects which influence the spectral shape need to be described with high precision.
As presented in this work, the shape of the response of SDDs to electrons depends on the entrance-window thickness and the backscattering probability.
To estimate the impact of an uncertainty of these properties, we performed a sensitivity study based on a semi-analytical detector model~\cite{Korzeczek.2020}. 

For the study, we assume a differential measurement with total statistics of $10^{18}$ electrons (corresponding to three years' data taking with KATRIN at a 100-fold reduced column density).
The detector response function is based on multiple interpolated MC simulations with KESS. The individual MC simulations were performed in a fine grid of various incident energies and incident angles. For this study, the detector response model takes into account the effect of the so-called post-acceleration electrode. This electrode, boosts the kinetic energy of all electrons by up to \SI{20}{\keV} on their way to the detector~\cite{Amsbaugh.2015}.\footnote{In the current KATRIN design the post-acceleration energy is limited to \SI{12}{\keV} for technical reasons.}

We investigate the impact of the uncertainty of a parameter of the response function (e.g.~the entrance-window thickness), by generating $10^3$ MC samples of the measured spectrum, each time varying the parameter of interest.
From these MC samples the variance of all data points and their covariance is deduced, i.e.~the covariance matrix $C$ is constructed.

The sensitivity of the experiment is derived by minimizing the squared data-to-model residuals for $40\times40$ grid points in the ($m_s$, $\sin^2(\Theta)$)-plane:
\begin{eqnarray}
\label{eq:chi2}
    \chi^2(m_s,\sin^2\Theta) = \vec{r}^{\mathsf{T}} C^{-1}\vec{r}~, \\
    \vec{r} \equiv \vec{r}(m_s,\sin^2\Theta) = \vec{R}_\mathrm{model}(m_s,\sin^2(\Theta))-\vec{R}_\mathrm{truth}(0,0)~,
\end{eqnarray}
where $\vec{R}_\mathrm{model}$ depicts the model expectation in case of a sterile neutrino with mass $m_s$ and mixing amplitude $\sin^2(\Theta)$, and $\vec{R}_\mathrm{truth}$ depicts the MC truth, for which no sterile neutrino is assumed.

At each grid point a minimization with respect to the spectrum normalization $N$, a constant background rate $B$ and the spectrum endpoint $E_0$ is performed.
The sensitivity at \SI{90}{\%} C.L.~is given by the contour of $\Delta\chi^2 = \chi^{2} - \chi^{2}_\mathrm{best} = 4.6$, where $\chi^{2}_\mathrm{best}$ is the $\chi^{2}$ value found for the case \mbox{($m_s$, $\sin^2(\Theta)$)} = ($0$, $0$).

\begin{figure}
    \centering
    \subfloat[Effect of uncertainty on the dead layer.]{
    	\includegraphics[width=0.45\textwidth]{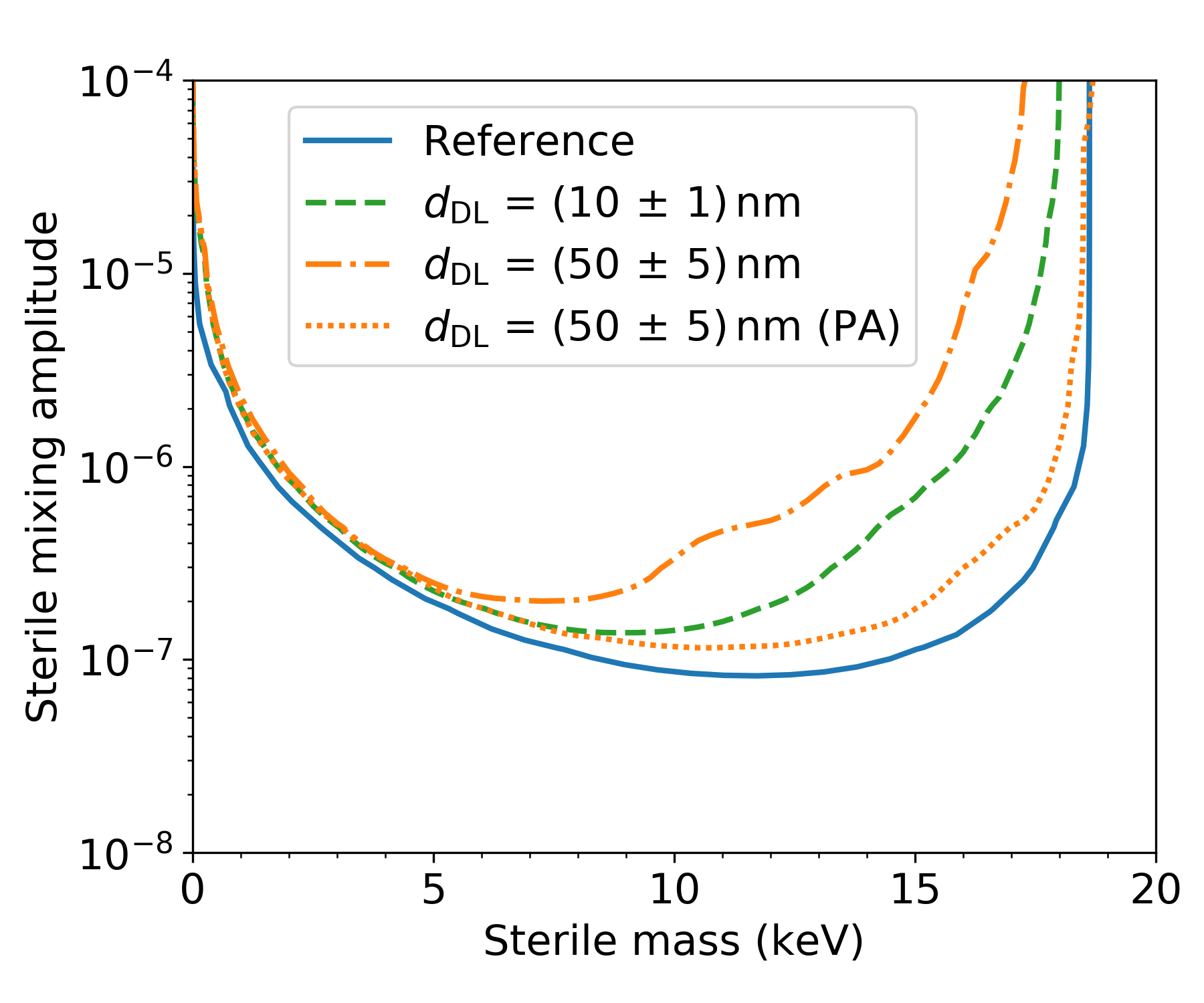}
    	\label{fig:DLBS:DL}
    }
    \subfloat[Effect of uncertainty on the incident angle.]{
    	\includegraphics[width=0.45\textwidth]{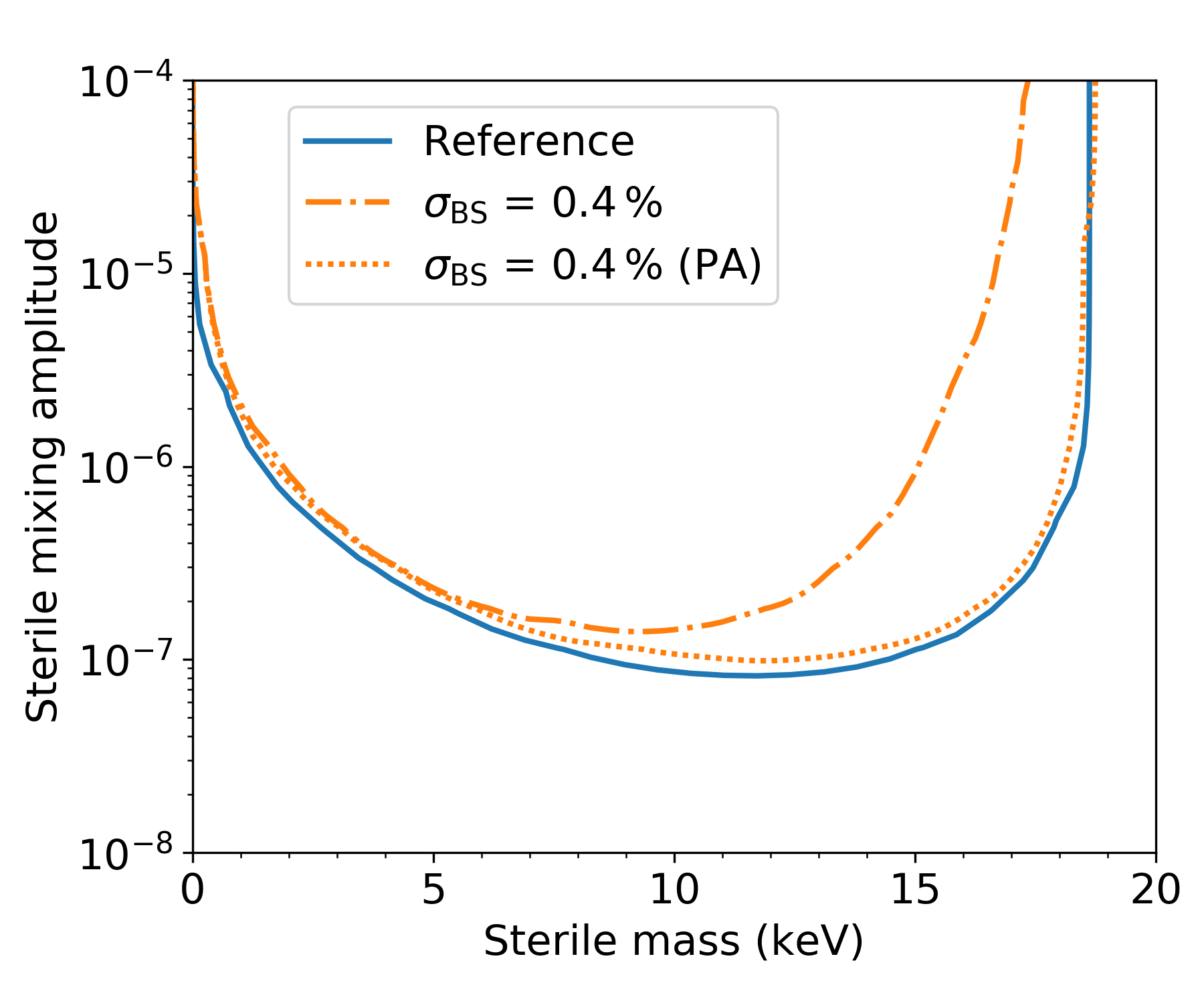}
   		\label{fig:DLBS:BS}
   	}
    \caption{Sensitivity study of the impact of detector uncertainty on the sterile-neutrino sensitivity. The solid blue line represents the sensitivity after three years' data taking with KATRIN at a \mbox{100-fold} reduced column density. No uncertainty on the dead-layer thickness or the incident angle is considered.
    \protect\subref{fig:DLBS:DL} In certain mass ranges, a \SI{10}{\%} uncertainty on a \SI{50}{\nm} thick dead layer reduces the sensitivity by a factor of ten (orange dash-dotted). Applying a post-acceleration (PA) energy of \SI{20}{\keV} almost fully recovers the sensitivity (orange dotted). The effect is significantly reduced for a smaller dead-layer thickness of \SI{10}{\nm} (green dashed).
    \protect\subref{fig:DLBS:BS} A \SI{0.4}{\%} uncertainty on the backscattering probability (\SI{5}{\degree} uncertainty on the electrons' incidence angle) significantly reduces the sensitivity of the experiment (orange dash-dotted). A post-acceleration can also mitigate this effect (orange dotted).}
    \label{fig:DLBS}
\end{figure}

In the first study we investigate the impact of an uncertainty on the entrance-window thickness of \SI{10}{\%}. As shown in fig.~\ref{fig:DLBS:DL}, a \SI{10}{\%} uncertainty on a \SI{50}{\nm} thick dead layer reduces the sensitivity by up to a factor of ten in certain mass ranges.
The effect is significantly reduced for a smaller dead-layer thickness of \SI{10}{\nm}.
Moreover, the effect can be almost fully eliminated by applying a post-acceleration energy of \SI{20}{\keV}, as the increased energy of the $\upbeta$-electrons reduces the relative fraction of energy loss in the dead layer.

In the second study we test the impact of an uncertainty of the backscattering probability. This uncertainty can for instance arise from an uncertainty of the incidence angle of the electrons, which itself could arise from an uncertainty of the exact orientation of the detector. For this study, we emulate this effect by assuming electrons with an incident angle of \SI{0 \pm 5}{\degree}. This translates to an uncertainty in the backscattering probability of \SI{0.4}{\%}~\cite{Renschler.2011} as well as in the experienced dead-layer thickness of \SI{0.38}{\%} (see fig.~\ref{fig:Tilt_Motivation}). 

The result, displayed in fig.~\ref{fig:DLBS:BS} shows, that an uncertainty on the backscattering probability of 0.4\% significantly reduces the sensitivity of the experiment. Again, the post-acceleration of electrons can fully mitigate the effect as high-energy electrons reach further into the detector and are less likely to backscatter.

\section{Conclusion and outlook}
\label{chap:conclusion}

The KATRIN experiment, equipped with a novel multi-pixel silicon drift detector (SDD) focal plane array, has the potential to perform a search for keV-scale sterile neutrinos with an unprecedented sensitivity, compared to previous laboratory experiments.
In the framework of the TRISTAN project, first SDD prototypes have been developed and an excellent performance with x-rays was demonstrated. 

In this work, mono-energetic electrons in the keV range from a scanning electron microscope and a radioactive \kr{} source were used to characterize the prototype detectors.
An excellent agreement between the observed electron spectra and an empirical analytical as well as a Monte-Carlo-based model was found. 

The entrance-window thickness, a key parameter of the detector response to electrons, was determined by tilting the detector and determining the relative shift of the main electron energy peak.
Assuming a step-like dead-layer model, a thickness of \SI{49\pm3}{\nm} was derived by comparing the measured energy shifts to Monte Carlo simulations.
With this result the requirements with respect to the energy resolution are met.

Finally, the impact of uncertainties of the SDD response to electrons on the final sterile-neutrino sensitivity was studied based on Monte Carlo simulations.
An entrance-window thickness of \SI{50\pm5}{\nm} and a backscattering uncertainty of 0.4\% decrease the sensitivity by about a factor of ten for sterile neutrino masses above \SI{10}{\keV}.
Reducing the entrance-window thickness and absolute uncertainty to \SI{10\pm1}{\nm} would mitigate this degradation.
As a major result, we find that the requirements on the uncertainty of dead-layer thickness and backscattering probability are reduced to an almost negligible level, when boosting the energy of the electrons with a post-acceleration electrode. 

The next stage of the TRISTAN project will be the integration of a SDD module with \mbox{166 channels} in the KATRIN monitor spectrometer in 2021~\cite{Erhard.2014}.
The final system, consisting of \mbox{21 modules}, will prospectively be integrated in the KATRIN beamline after successful completion of the neutrino mass measurement.

\ack
We acknowledge the support of Helmholtz Association (HGF), Ministry for Education and Research BMBF (05A17VK2 and 05A17WO3), the doctoral school KSETA at KIT, the Max Planck Research Group (MaxPlanck@TUM) program, and the Deutsche Forschungsgemeinschaft DFG (GSC-1085-KSETA and SFB-1258).
This project has received funding from the European Research Council (ERC) under the European Union Horizon 2020 research and innovation program (grant agreement No. 852845).
This work is further supported by the U.S. Department of Energy, Office of Science, Office of Nuclear Physics under Award Number DE-AC05-00OR22725, by the Ministry of Education, Youth and Sport (CANAM-LM2015056, LTT19005), and by the Ministry of Science and Higher Education of the  Russian Federation under the contract 075-15-2020-778.

\printbibliography

\end{document}